\documentclass{egpubl}
\usepackage{eurovis2026}

\SpecialIssuePaper         

\CGFccby

\usepackage[T1]{fontenc}
\usepackage{dfadobe}  

\usepackage{cite}  
\BibtexOrBiblatex
\electronicVersion
\PrintedOrElectronic
\ifpdf \usepackage[pdftex]{graphicx} \pdfcompresslevel=9
\else \usepackage[dvips]{graphicx} \fi

\newcommand{\name}{\textit{SemiConLens}}

\newcommand{\inlineicon}[1]{\raisebox{-0.3\baselineskip}{\includegraphics[height=\baselineskip]{figs/#1}}}

\graphicspath{{src/}{figs/}} 

\usepackage{amsmath} 
\usepackage{amssymb} 
\usepackage{svg}
\usepackage{makecell} 
\usepackage{multirow}  
\usepackage{tabu}                      
\usepackage{booktabs}                  
\usepackage{caption}

\usepackage{algpseudocode} 
\usepackage{algorithm} 
\algrenewcommand\alglinenumber[1]{\hspace*{-\algorithmicindent}#1:}
\algrenewcommand\algorithmicindent{0.8em}

\usepackage{egweblnk}

\title[SemiConLens: Visual Analytics for 2D Semiconductor Discovery]%
      {SemiConLens: Visual Analytics for 2D Semiconductor Discovery}

\author[K. Athapaththu, S. Chen, F. Yuan, S. Mitra, Y. S. Ang \& Y. Wang]
{\parbox{\textwidth}{\centering 
        Kavinda Athapaththu$^{1,2}$\orcid{0000-0002-8641-7768},
        Shiwei Chen$^{1}$\orcid{0009-0007-9106-7989},
        Yuan Fang$^{2}$\orcid{0000-0002-4265-5289},
        Sanchali Mitra$^{3}$\orcid{0000-0002-3829-0010},
        Yee Sin Ang$^{3}$\orcid{0000-0002-1637-1610},
        Yong Wang$^{1,2}$\thanks{Corresponding author: yong-wang@ntu.edu.sg}\orcid{0000-0002-0092-0793}
        }
        \\
{\parbox{\textwidth}{\centering $^1$Nanyang Technological University\\
         $^2$Singapore Management University\\
         $^3$Singapore University of Technology and Design\\
       }
}
}

\begin{document}


\maketitle
\begin{abstract}
   The past few years have witnessed vibrant efforts in discovering new two-dimensional (2D) semiconductor materials from both academia and the industry, due to their promising potential in resolving the severe performance deterioration of traditional semiconductors resulting from condensed silicon thickness.
    However, existing methods (e.g., Density Functional Theory (DFT) or machine-learning-based approaches) suffer from various challenges such as small datasets, and reliability and trustworthiness issues.
    To bridge this gap, we propose \name, a visual analytics approach to combine human expertise with the power of ML to enable effective and reliable 2D semiconductor discovery.
    Specifically, we first develop a new Correlation Aware
Multivariate Imputation (CAMI) method and use ML models like autoencoder, which can better learn from limited data and reveal uncertainty, to address the challenge of sparse data in semiconductivity prediction. Built upon this, our visualization module, consisting of three visualization views with linked interactions, allows material researchers to interactively filter, discover and compare 2D semiconductor candidates. A novel circular glyph design and a new cluster-aware layout optimization approach are proposed to effectively display all the user-configurable key attributes and possible prediction uncertainties of each semiconductor candidate, ensuring a reliable and trustable 2D semiconductor discovery.
    We assess \name\ through quantitative evaluations, expert interviews, and use cases. The results demonstrate \name's capability to help material researchers conduct effective discovery of desirable 2D semiconductors.

\ccsdesc[500]{Human-centered computing~Visualization systems and tools}

\printccsdesc   
\end{abstract}  
\section{Introduction}\label{section::Intro}

Semiconductors are essential to modern technologies in areas such as quantum computing, communication, health care, energy, transportation, and defense, with the global market reaching 588.4 billion USD in 2024 and projected to grow to 654.7 billion USD in 2025 \cite{sia_factbook_2023,sia_factbook_2024}. A major challenge is the performance degradation of silicon transistors at nanometer-scale thicknesses \cite{nano_meter_10_ieong_silicon_2004,2d_semi_extend_moores_li20192d}. The discovery of graphene in 2004 \cite{graphene_Novoselov} showed that two-dimensional semiconductors, semicondcutors only a few atoms thick, can overcome these limits and offer superior performance \cite{2d_semi_extend_moores_li20192d}. As a result, discovering new 2D semiconductor materials with desirable properties has become a highly active research area \cite{c2db_gjerding2021recent,eTran2d,highthroughput_luo2021}, with major industry players such as TSMC, Intel, Nvidia, IBM, and Samsung investing heavily in this direction \cite{industry_world_trade_organization_global_2023}.

However, 2D semiconductor discovery is inherently a multi-criteria and high-dimensional problem \cite{materialproject_jain2013commentary}. Existing methods for 2D semiconductor material discovery can be categorized into two groups:
\textit{traditional experimental approaches}~\cite{butler2018machine,oliynyk2018discovery,Accelerating_materialscorrea2018} and \textit{computational approaches}~\cite{highthroughput_luo2021,c2db_gjerding2021recent}.

Traditional experimental workflows depend on researchers synthesizing candidate materials based on domain knowledge and then measuring their properties to evaluate suitability. Because this process is fundamentally trial-and-error, it is slow and requires extensive manual effort \cite{ML_Accelerating_chen2020machine,Accelerating_materialscorrea2018,MD_using_ML_liu2017materials,matexplorer_pu2021}.
To mitigate this issue, materials science researchers have also developed a series of computational approaches for 2D material discovery, which leverage theoretical models, simulations, and Machine Learning (ML) techniques to predict new 2D materials with desirable properties before experimental synthesis~\cite{c2db_gjerding2021recent,2dmatpedia_zhou2019,interpretabledis_choubisa2021,highthroughput_luo2021,butler2018machine,reliableandex_kailkhura2019}.
Representative computational approaches include: 1) using Density Functional Theory (DFT) for calculating material attributes, and 2) employing ML to predict 2D material attributes by training on existing material data.
DFT is a quantum mechanical simulation technique that can predict the electronic, optical and structural attributes of materials. However, with increasing material size, the methods become \textit{computationally expensive and slow}~\cite{highthroughput_luo2021,eTran2d}.

ML-based approaches have been increasingly popular for 2D semiconductor attribute prediction~\cite{Combinatorial_screening_meredig2014,butler2018machine,mlforsemi_LIU2022100033}, but  
ML-based approaches often require a large amount of labeled data to train the ML models to be effective. 

A fundamental challenge in 2D material discovery is \textbf{data sparsity}, as materials science datasets often contain \textbf{
limited labeled samples
}~\cite{schmidt2024improving,butler2018machine,zhang2018strategy,chang2022towards}, restricting the prediction performance of ML-based approaches.
For example, the C2DB dataset~\cite{c2db_gjerding2021recent}, one of the most popular 2D material datasets, has a total number of 16,905 compounds, but there are only 339 compounds with a given value for CBM GW and VBM GW, two important attributes to indicate the efficiency of semiconducting material in high-speed electronic devices.
Also, existing machine learning approaches, especially the deep neural network models, often work like a black box, negatively affecting the \textbf{reliability and trustworthiness} of these approaches~\cite{meng2024deep}.

To address the above challenges, we propose \name, a novel visual analytics approach to help material researchers with effective 2D semiconductor discovery by seamlessly combining the power of ML with their materials science expertise.
\name\ consists of two modules: \textbf{sparsity-aware prediction} and \textbf{visualization}.
Our semiconductivity prediction module explicitly considers the sparsity of material data labels by developing a Correlation Aware Multivariate Imputation (CAMI) method to fill the missing attribute values.

Built upon this, our visualization module allows material researchers to interactively analyze a large number of 2D compound materials, and easily compare and identify high-potential semiconductors. The system comprises three coordinated views: \textit{Filter View} to display the compound attribute distribution and user exploration history; \textit{Discovery View} to visualize filtered compounds and the similarities among them; and \textit{Comparison View} to facilitate detailed comparison and investigation of selected 2D semiconductor candidates.
Specifically, we propose a novel circular glyph to encode material attributes together with associated prediction uncertainties, and a new cluster-aware layout optimization technique to preserve inter and intra cluster similarity while improving readability in dense layouts. 
Together, these components support informed interactive exploration and transparent decision making in 2D semiconductor discovery. We evaluate the effectiveness and usability of \name\ through quantitative metric evaluations, expert interviews, and two use cases.
The major contributions of this paper are as follows:
\begin{itemize}
    \item We develop a new Correlation Aware Multivariate Imputation (CAMI) technique to impute the missing attribute values in the material dataset, reducing the influence of data sparsity challenge in the materials science field.

    \item We propose \name, a visual analytics approach for 2D semiconductor discovery, which integrates sparsity-aware prediction and interactive visualizations for effective and trustworthy 2D semiconductor discovery. A novel circular glyph design and cluster-aware glyph layout optimization are presented.
    \item We evaluate \name\ via quantitative metric evaluations, expert interviews, and use cases. The results demonstrate that \name\ is both effective and user-friendly for aiding material experts in discovering high-potential 2D semiconductors.
    
\end{itemize}

\section{Background}

In this section, we introduce materials science background knowledge essential to understand this research including desirable semiconductor attributes and different methods to calculate the above.
\subsection{Desirable Semiconductor Properties} \label{sec:desirable_semiconductor_propoerties}

\textit{Carrier Mobility} shows how fast an electron (\textit{Electron Mobility}) or a hole (\textit{Hole Mobility}) can travel in a semiconductor when an external electric field is applied. It determines properties of semiconductor devices like switching frequency in transistors~\cite{carrier_mob_ponce2020first}.

For any semiconductor, there is an energy region where there can be no states. The energy bands below are known as Valence Bands while energy bands above are known as Conduction Bands. The difference between the lowest conduction band, which is the \textit{Conduction Band Minimum (CBM)} and the highest valence band, which is the \textit{Valence Band Maximum (VBM)}, is known as the \textit{Bandgap}~\cite{bandgap_sze2021physics}. Electrons in the valence band are bound to atoms while the electrons in the conduction band are free to conduct electricity~\cite{bandgap_streetman2000solid}. Suitable bandgap, high carrier mobility and stability at room temperature are the desirable properties of semiconductors~\cite{eTran2d}. 

\subsection{Density Functional Theory (DFT) and beyond DFT}

DFT is a widely used computational method for studying electron distribution in materials. Rather than solving the full Schrödinger equation for every electron, it models the overall electron density, making calculations more feasible while remaining accurate for many ground-state properties~\cite{DFT_martin2020electronic}. Using DFT, semiconductor properties such as carrier mobility and bandgap can be estimated, but the computations are both expensive and time consuming.

Within DFT, the \textbf{Perdew–Burke–Ernzerhof (PBE) }functional~\cite{PBE_perdew_generalized_1996} is widely used, improving upon earlier methods by accounting for spatial variations in electron density and yielding more realistic results than the local density approximation. The \textbf{Heyd–Scuseria–Ernzerhof (HSE) }functional~\cite{hse_heyd_hybrid_2003} further enhances accuracy by mixing a portion of exact exchange from Hartree–Fock theory with DFT, allowing short- and long-range electron interactions to be treated differently. For even higher accuracy, particularly for electronic excitations, the \textbf{GW approximation}~\cite{gw_shi_optical_2010} extends beyond standard DFT; here, “G” describes electron or hole motion while “W” captures their interactions, producing results that often match experimental measurements more closely than PBE or HSE~\cite{gw_shi_optical_2010,hse_heyd_hybrid_2003,PBE_perdew_generalized_1996}.

\section{Related Work}

Our work is related to prior research on ML for material discovery, visualization for material discovery, and high-dimensional data visualization.

\subsection{ML for Material Discovery} 

ML has accelerated materials discovery by enabling rapid screening and prediction of novel compounds. ML approaches are commonly categorized by their reliance on labeled data, as supervised, semi-supervised, and unsupervised learning~\cite{ML_Accelerating_chen2020machine}, or by the primary data mining task such as dimension reduction, clustering, classification, and correlation~\cite{SOA_in_interg_endert2017}. Among these, supervised learning is most widely used for predicting material properties, identifying stable compounds, and designing new materials~\cite{ML_Accelerating_chen2020machine,MD_using_ML_liu2017materials}.

Supervised algorithms including Random Forest (RF) ~\cite{mlforsemi_LIU2022100033}, Support Vector Regression (SVR) ~\cite{ML_Accelerating_chen2020machine}, Multi-Layer Perceptron (MLP) ~\cite{lu2018accelerated}, Decision Tree (DT) ~\cite{eTran2d,lu2018accelerated}, and Gradient Boost (GB) ~\cite{reliableandex_kailkhura2019,lu2018accelerated} have been applied to tasks such as predicting ionic conductivity ~\cite{matexplorer_pu2021}, designing lead-free perovskites ~\cite{lu2018accelerated}, and classifying crystal structures~\cite{oliynyk2018discovery}. Ensemble methods like GB are particularly effective for uncovering nonlinear relationships in high-dimensional datasets~\cite{reliableandex_kailkhura2019}. However, no single model consistently outperforms others due to the complexity of materials data~\cite{ML_Accelerating_chen2020machine}.

Unsupervised and semi-supervised methods help leverage large unlabeled datasets in materials discovery. Techniques like k-means and t-SNE group materials with similar properties \cite{c2db_gjerding2021recent}, while Auto Encoders (AE) and other neural networks have shown advantages in extracting complex structures from high-throughput materials datasets and enabling accelerated screening of candidate materials\cite{AE_for_Material_Scienceyamaguchi_drawing_2023,highthroughput_luo2021}. Semi-supervised methods further combine small labeled datasets with abundant unlabeled data to predict properties such as phase stability and thermodynamics \cite{interpretabledis_choubisa2021}.

Although previous studies have explored various applications of machine learning in the materials science domain, none have been specifically tailored for the discovery of 2D semiconductors. Furthermore, to the best of our knowledge, the use of few-shot learning with autoencoders has not been investigated in this context.

\subsection{Visualization for Material Discovery}

Visualization offers valuable support in materials discovery, where researchers must navigate complex, high-dimensional datasets to identify promising candidates~\cite{SOA_in_interg_endert2017, FS_based_on_VA_for_quality}. While still emerging in this domain, visual analytics approaches have shown potential to enhance the exploration and analysis of material properties \cite{design_of_new_dispersants,phasemapper_bai2018,matexplorer_pu2021}.

Early visualization systems mainly supported basic property inspection and relationship discovery. Martínez et al.~\cite{design_of_new_dispersants} employed scatter plots and scatter matrices, but these techniques were insufficient for high-dimensional materials data. Visualization also assisted model interpretation, as seen in Phase Mapper~\cite{phasemapper_bai2018}, which connected model predictions with expert assessment using simple visual tools. MatExplorer~\cite{matexplorer_pu2021} further integrated predictive modeling with interactive exploration for solid-state electrolytes. However, it remains inadequate for semiconductor discovery because it does not enable progressive exploration of large-scale datasets and lacks support for in-depth examination of influential attributes.

To overcome these limitations, this paper introduces a visual analytics system that enables systematic exploration of 2D semiconductor materials through interactive visualizations, progressive filtering, and uncertainty representation. The system incorporates a glyph-based design for examining semiconductor properties and a cluster-aware layout optimization method that combines cartograms with force-directed layouts to minimize overlap while preserving cluster structure.

\begin{figure}[h]
  \centering
  \includegraphics[width=\columnwidth]{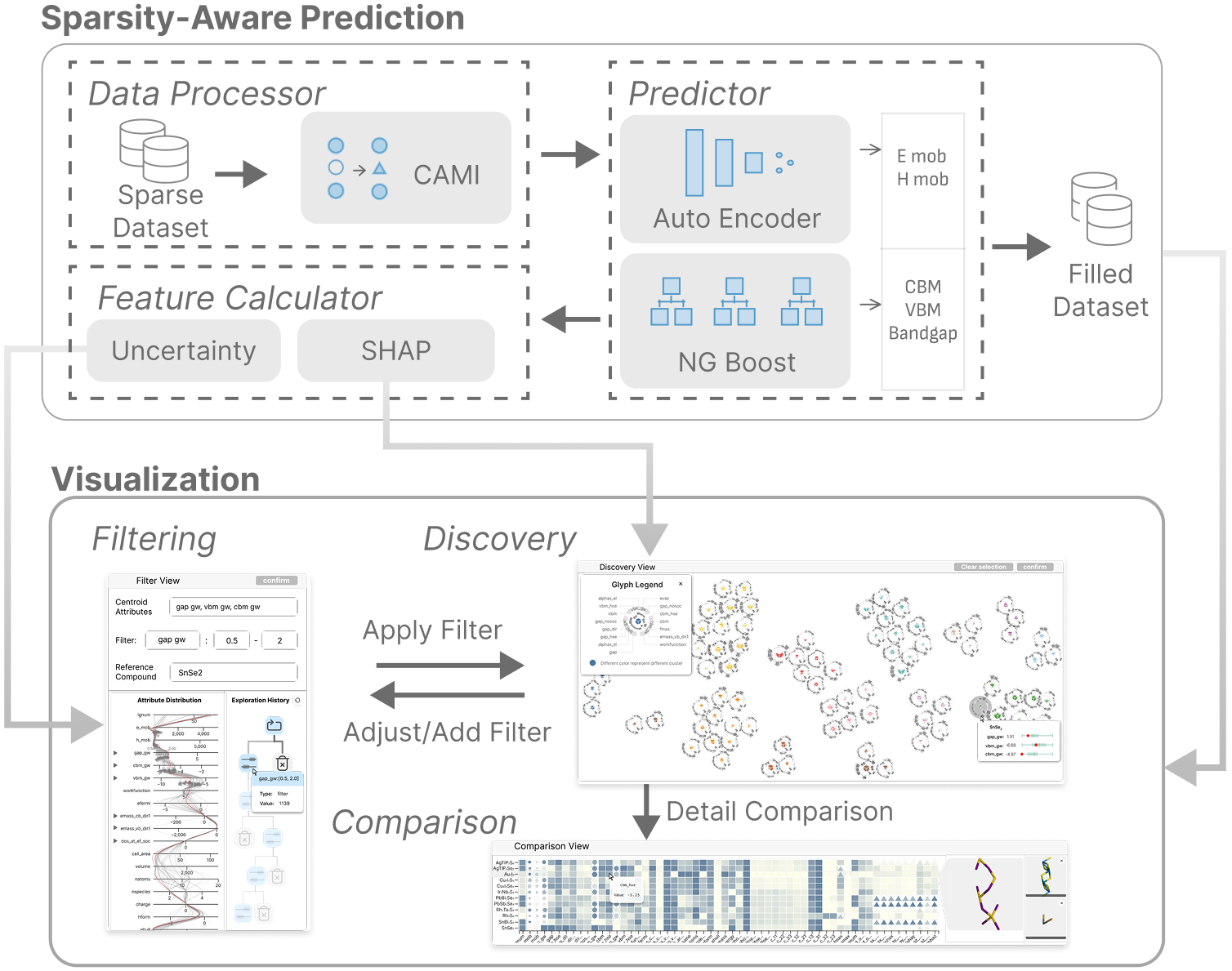}
  \caption{The workflow consists of two main components: (1) a sparsity-aware prediction pipeline that processes sparse datasets through CAMI data processing, followed by prediction using Auto Encoder and NG Boost models, with uncertainty and SHAP-based feature calculation for interpretability; (2) a visual analysis system, which enables interactive \textit{filtering}, \textit{exploration} and \textit{comparison} of the filled dataset.}
  \label{fig:overview}
\end{figure}

\subsection{High Dimensional Data Visualization}

High-dimensional data visualizations aim to improve the interpretability of complex multivariate datasets. Among them, glyph-based approaches are mainstream techniques, offering effective information preservation while reducing cognitive load.

For glyph design, multiple dimensions are encoded into a single visual object using properties such as color, shape, and size~\cite{chen2008multivariate, borgo2013glyph, SOA_in_interg_endert2017,ChemVA_sabando_chemva_2020}. Different designs serve specific analytical purposes: star plots highlight extremal values, Chernoff faces exploit humans’ sensitivity to facial feature differences, and radial encodings represent cyclical or hierarchical relationships~\cite{fuchs2013evaluation}. Inspired by existing glyph designs, we adopted a radial encoding to illustrate key attributes and corresponding influencing factors in our visualizations.

For glyph layout, Dimension Reduction (DR) techniques are widely applied to project high-dimensional data into lower-dimensional spaces while preserving key relationships~\cite{blumberg5108314multiinv}. Common DR methods such as PCA~\cite{SOA_in_interg_endert2017}, t-SNE~\cite{reliableandex_kailkhura2019}, and UMAP~\cite{ML_Accelerating_chen2020machine}
are widely used for glyph layout.
However, DR approaches treat data points as dimensionless, ignoring the spatial requirements of glyphs, which causes overlap in dense regions, particularly for large datasets. Existing solutions such as force-directed layouts~\cite{hu2005efficient}, constraint-based techniques~\cite{zhou2024ClusterAware}, and space partitioning methods~\cite{hilasaca2024GridBased} primarily address overlap reduction but do not explicitly preserve cluster structures in DR projections. This paper presents a novel cluster-aware glyph layout optimization approach to address this challenge.

\section{Tasks Analysis}
\label{sec::task analysis}
Over more than 12 months, we worked closely with two domain experts, who are also co-authors, to identify challenges in 2D semiconductor discovery and iteratively refine our prototypes. Their continued feedback allowed us to distill six core analytical tasks that shaped the system’s design.

\begin{description}
    \item \textbf{T1. Identify high-potential models to predict compound attributes that are hard to calculate using DFT or other traditional methods.}
    Traditional methods such as DFT struggle to compute several semiconductor attributes efficiently~\cite{highthroughput_luo2021,eTran2d}. Thus, researchers have explored multiple ML models to predict these properties~\cite{Combinatorial_screening_meredig2014}. Since model performance varies across datasets and target attributes~\cite{matexplorer_pu2021}, it is essential to determine which ML models offer the highest predictive potential for these attributes.

    \item \textbf{T2. Visualize the uncertainty and rationales of the predictions.}
    ML-based predictions inherently involve uncertainty, making it essential for material scientists to understand the reliability of the results~\cite{reliableandex_kailkhura2019,interpretabledis_choubisa2021}. This requires examining both the uncertainty in the predicted values and the underlying rationale for the model’s decisions. Providing clear insight into model confidence and reasoning enables users to make informed judgments when interpreting prediction outcomes.
    
    \item \textbf{T3. Select key attributes flexibly and show influencing factors.}
    Material scientists often prioritize different attributes depending on their application needs~\cite{matexplorer_pu2021,highthroughput_luo2021,kahle2020high}. Thus, enabling flexible selection and visualization of relevant attributes is essential. Moreover, examining relationships among these attributes helps identify key factors that influence material performance, supporting deeper analysis and more informed decision-making.
    
    \item \textbf{T4. Provide an overview of the attributes distributions.} 
    Understanding how attributes are distributed across compounds is vital for material scientists~\cite{matexplorer_pu2021}. Distribution overviews enable the identification of outliers and provide insight into dataset-wide patterns. Clear visualizations of these distributions help users assess ranges, frequencies, and central tendencies, offering context for evaluating individual compounds. Such insights support the detection of promising regions in the material space and facilitate the development of data-driven hypotheses for further design and investigation.

    \item \textbf{T5. Enable progressive filtering exploration.}
    For large datasets, material scientists must progressively narrow the search space through systematic compound filtering~\cite{matexplorer_pu2021,phasemapper_bai2018}. When clear criteria are available, users should be able to iteratively apply and refine multiple filter parameters. When such criteria are absent, the system should support non-parametric exploration and guide users by suggesting potential filtering strategies informed by attribute distributions and statistical patterns derived from exploratory analyses.
  
    \item \textbf{T6. Compare possible candidate compounds.}
    After narrowing the search space, material scientists must compare candidate compounds across multiple attributes. The system should support both overall evaluation to determine the best-performing compound and detailed, attribute-level comparisons to highlight individual strengths and weaknesses.
\end{description}

\section{Sparsity-aware Prediction}
This section summarizes our approach for predicting compound attributes, covering model selection, missing-value imputation, prediction and uncertainty estimation, and the data sources and processing procedures.

\subsection{Correlation Aware Multivariate Imputation}

The correlation between compound attributes and the similarity across different compound samples can facilitate the estimation of missing compound attributes, which, however, is not fully considered in by existing data imputation techniques. We propose a novel imputation method, Correlation Aware Multivariate Imputation (CAMI), which considers the above to iteratively impute the missing values. A sketch of the above method is as described in Algorithm 1 in Appendix A.7. Fig. \ref{fig:caimi_overview} demonstrates a summarized example of the behavior of the CAMI algorithm (See Appendix A.6 for high resolution version). We used CAMI to impute missing data in the dataset, except for the desirable semiconductor properties discussed in section \ref{sec:desirable_semiconductor_propoerties}, which are predicted.

CAMI begins with a preprocessing step (Fig.  \ref{fig:caimi_overview}\inlineicon{caimi_overview_icons/A2.png}) where it selects columns that have less than a specified threshold $A\%$ of missing values and drops any columns exceeding this threshold to prevent noise from features that are largely incomplete, and the dropped columns are not imputed and will not be used in the system . Then, rows with less than $B\%$ missing values are selected for the same reason, removing and never filling rows that exceed this threshold. After preprocessing,  CAMI computes the correlation matrix ($c_m$) using Pearson Correlation with pairwise complete observations , on the remaining data (Fig.  \ref{fig:caimi_overview}\inlineicon{caimi_overview_icons/A3.png}). For each target column $Y$ containing missing values, the algorithm identifies the top $C$ most correlated columns ($K$) using $c_m$ (Fig.  \ref{fig:caimi_overview}\inlineicon{caimi_overview_icons/B1.png}). The missing values in these selected columns are then imputed using the mean of the observed values in each respective column (Fig.  \ref{fig:caimi_overview}\inlineicon{caimi_overview_icons/B2.png}).

Then, for each target column Y, the algorithm employs a similarity-based row selection and imputation. First, from $X$, the most correlated columns ($K$) and the column $Y$ are selected ($X'$) (Fig.  \ref{fig:caimi_overview}\inlineicon{caimi_overview_icons/B3.png}) and the subset of the dataset ($X'$) is split into two subsets: $L$, containing rows where $Y$ is missing, and $M$, containing rows where $Y$ is observed (Fig.  \ref{fig:caimi_overview}\inlineicon{caimi_overview_icons/B4.png}). For each row $P$ in $L$, the similarity to every row in $M$ is computed based on the available features, using K-Nearest Neighbors (Fig.  \ref{fig:caimi_overview}\inlineicon{caimi_overview_icons/C1.png}). The algorithm then selects the top $D$ most similar rows ($O$) from $M$ (Fig.  \ref{fig:caimi_overview}\inlineicon{caimi_overview_icons/C2.png}) and estimates the missing value of $Y$ in $P$ using the mean value of $Y$ from the selected similar rows (Fig.  \ref{fig:caimi_overview}\inlineicon{caimi_overview_icons/C3.png}). This  ensures that missing values are imputed using contextually relevant instances, rather than relying solely on a global mean. However, this estimated value will not be used for subsequent similarity searches to prevent compounding errors. The similarity-based row
selection and imputation is repeated to fill all the missing values from remaining columns and rows. The hyperparameters were tuned specifically for the dataset through an iterative exhaustive search procedure.

\begin{figure}[bt]
  \centering
  \includegraphics[width=\columnwidth]{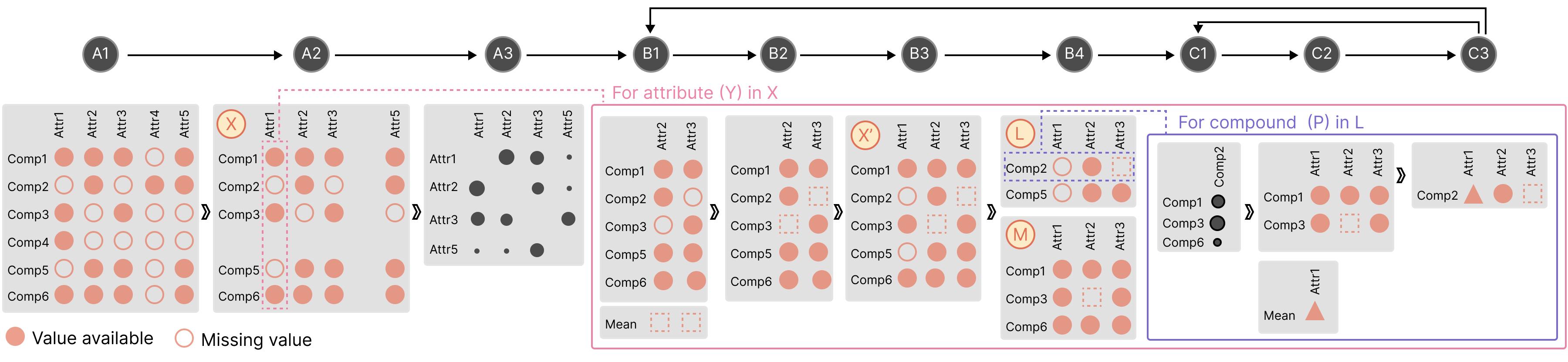}
  \caption{An illustration of the proposed Correlation Aware Multivariate Imputation. 
  Step \inlineicon{caimi_overview_icons/A1.png} : Load the dataset. 
  Step \inlineicon{caimi_overview_icons/A2.png} : Select only columns and rows with less than $A\%$ and $B\%$ missing values respectively.
  Step \inlineicon{caimi_overview_icons/A3.png} :  Calculate the correlation matrix.
  Step \inlineicon{caimi_overview_icons/B1.png} : For each attribute ($Y$) in the filtered dataset $X$, select top $C$ most correlated columns  using the correlation matrix. Then calculate the mean for each selected column.
  Step \inlineicon{caimi_overview_icons/B2.png} : Fill the missing values of each column with the corresponding calculated mean.
  Step \inlineicon{caimi_overview_icons/B3.png} : Create $X'$, a subset of $X$, using the $Y$ and its most correlated columns mentioned above. 
  Step \inlineicon{caimi_overview_icons/B4.png} : split $X'$ into $L$ and $M$, where column $Y$ is missing and not missing respectively. 
  Step \inlineicon{caimi_overview_icons/C1.png}: For each compound ($P$) in $L$, compute similarity matrix with compounds in $M$. 
  Step \inlineicon{caimi_overview_icons/C2.png}: Select the top $D$ most similar compounds from the subset and calculate the mean for column $Y$.
  \inlineicon{caimi_overview_icons/C3.png} : Fill the missing value of $P$ with the mean calculated above.
  }
\label{fig:caimi_overview} 
  
\end{figure}

\subsection{Model Choice}
Based on the comparative evaluation results discussed in Table \ref{tab:model_comparison}, AE was selected for predicting electron mobility and hole mobility, and Natural Gradient Boosting (NGBoost) model for CBM GW, VBM GW and Bandgap GW (\textbf{T1}). This decision was supported by the fact that the authors of NGBoost successfully tested the datasets with as few as 309 data points~\cite{NGBoost_duan_ngboost_2020}.

 AE is a specialized neural network trained to compress and then reconstruct the inputs~\cite{AutoEncoder}. 
 It captures underlying patterns from large unlabeled datasets and also supports few-shot learning for predicting attributes when limited data is available~\cite{ae_for_few_shot_bindini_tiny_2024}. NGBoost is a probabilistic prediction algorithm that extends gradient boosting to model full probability distributions rather than single point estimates, demonstrating strong performance on smaller datasets~\cite{NGBoost_duan_ngboost_2020}. Both AE and NGBoost inherently provide uncertainty measures: AE through reconstruction error and NGBoost through predicted standard deviation. All models were trained using the preprocessed dataset, where missing values were imputed using the CAMI method.

\subsection{Uncertainty and Feature Importance Estimation}
\label{sec::explainable predictions}

As discussed in the introduction, the reliability and trustworthiness of ML models remain a concern. During the task analysis, we also identified the need to visualize uncertainty and the rationale behind predictions (\textbf{T2}). To support this, we compute both the prediction uncertainty and the feature importance for each model output.

\textbf{Predicting semiconductor attributes.}
According to the previous results, two AE models were fine-tuned for the prediction of electron mobility and hole mobility. Bandgap GW, VBM GW, and CBM GW were predicted using three natural gradient boosting models.

\textbf{Calculating uncertainty.}
Uncertainty was computed in two ways. For AE based predictions, we use the reconstruction error prior to finetuning as the measure of uncertainty\cite{uncertanity_AE_pimentel_review_2014}. This error indicates how well the model reproduces the original input from its latent representation. Given an input vector
\(\mathbf{x} = (x_1, x_2, \ldots, x_n)\) 
and its reconstruction using the AE model
\(\hat{\mathbf{x}} = (\hat{x}_1, \hat{x}_2, \ldots, \hat{x}_n)\), 
the reconstruction error using Mean Absolute Percentage Error (MAPE) is defined as 
\( \text{MAPE} = \text{Reconstruction error }(\mathbf{x}, \hat{\mathbf{x}}) 
= \frac{1}{n} \sum_{i=1}^{n} \left| \frac{x_i - \hat{x}_i}{x_i} \right| \times 100\% .\)

For natural gradient boosting models, we leverage Relative Standard Deviation (RSD) also known as the coefficient of variation, to calculate uncertainty of the predictions ~\cite{chen_prediction_2019_RSD,addepalli_quantifying_2021_RSD}. The NGB model inherently provides both the mean ($\mu$) and standard deviation ($\sigma$) as outputs, representing the central estimate and associated uncertainty, respectively. The RSD can be defined as \( \text{RSD} = \left( \frac{\sigma}{|\mu|} \right) \times 100\% \).

The uncertainty is not estimated for the imputed values, but they are clearly marked as imputed values in the visualization.

\textbf{Calculating feature importance.}
To interpret model predictions and understand which features contribute most, we use SHAP (SHapley Additive exPlanations)~\cite{SHAP}, a game-theoretic approach to explain the output of machine learning models. First, local feature importance, specific to each compound, is calculated using SHAP values. Global feature importance is then derived by averaging these SHAP values across all compounds. This process is repeated for each of the five models. Detailed discussion about data sources and preprocessing can be found in Appendix A.2.

\begin{figure*}[bt]
  \centering
    \setlength{\abovecaptionskip}{0.2cm}
  \setlength{\belowcaptionskip}{-0.4cm}
  \includegraphics[width=0.99\textwidth]{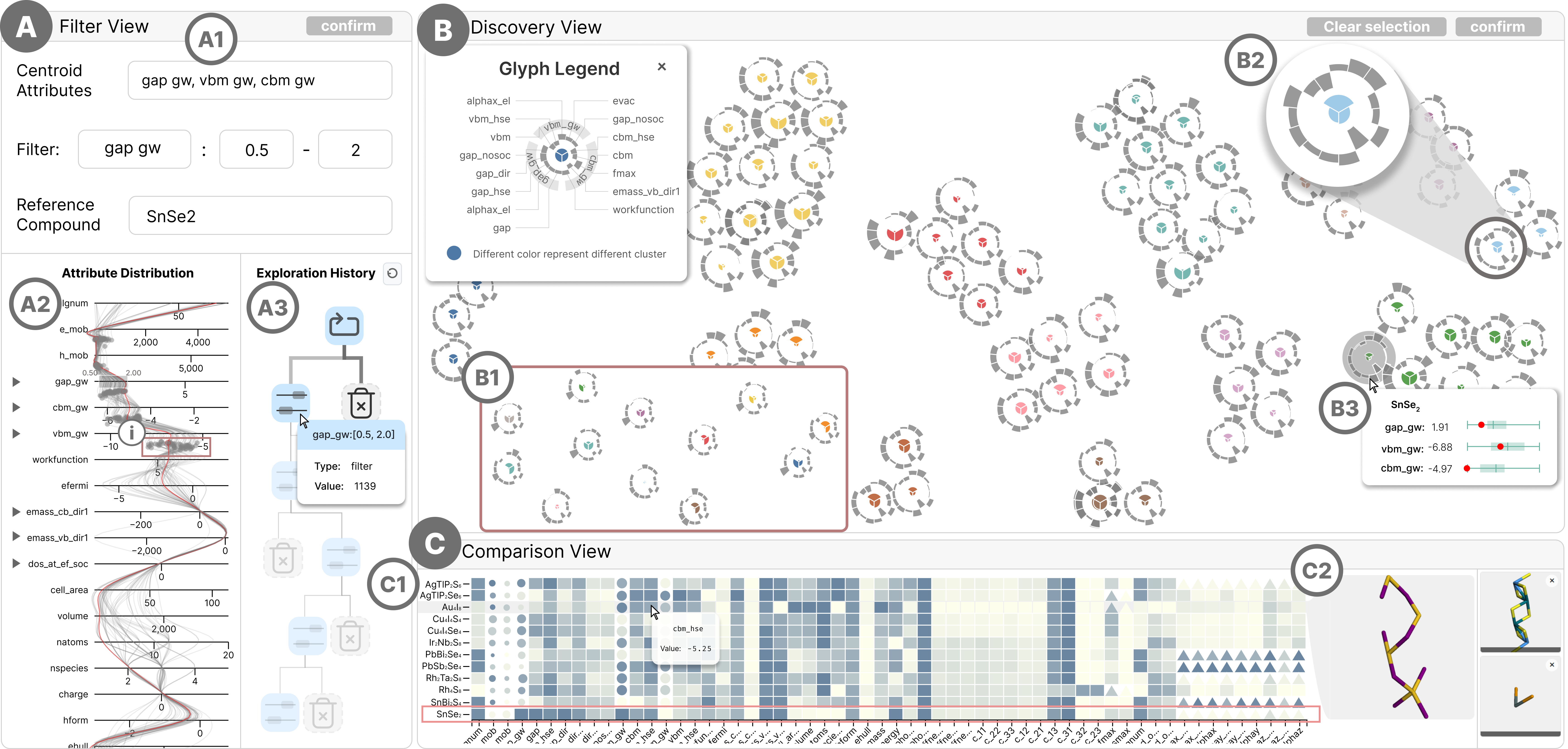}
  \caption{System interface for \textit{SemiConLens}. Filter View \inlineicon{case_2_icons/A.png} provides an overview of the compound space, applies range-based filters, and tracks exploration history for a more targeted and efficient analysis. Discovery View \inlineicon{case_2_icons/B.png} visualizes clusters and influencing factors of the key attributes, enabling users to discover patterns in the compound space. Comparison View \inlineicon{case_2_icons/C.png} facilitates the comparison of compounds across all attributes and displays their structures. See Appendix A.8 for high resolution version.
  }
  \label{fig:teaser}
\end{figure*}

\section{Visual Design}
Based on the design tasks in section \ref{sec::task analysis}, we developed \name\ (Fig.\ref{fig:teaser}), an interactive visual analytics system with three coordinated views: Filter View, Discovery View, and Comparison View. Fig.\ref{fig:overview} showcases the overall workflow of our system (See Appendix A.3 for high resolution version). In this section, we present detailed descriptions of each view's design and interactions . A detailed discussion on interaction design can be found in Appendix A.1.

\subsection{Filter View}\label{section::FilterView}

The Filter View (Fig. \ref{fig:teaser}\inlineicon{case_2_icons/A.png}) is designed to enable flexible exploration, and provide an overview of the compound space (\textbf{T3, T4, T5}). The Filter View comprises three sub-views: Control Panel, Attribute Distribution, and Exploration History.
\par
\textbf{Control Panel }(Fig. \ref{fig:teaser}\inlineicon{case_2_icons/A1.png}) serves as an interactive interface for attribute selection, filtering configuration, and reference semiconductor selection. It consists of three main components: (1) an attribute selector that enables users to choose multiple key attributes for detailed analysis in the Discovery View (\textbf{T3}), (2) a filtering criteria configuration panel that provides precise controls for configuring range filtering (\textbf{T5}), and (3) a reference semiconductor selector that allows users to incorporate known semiconductor materials as benchmarks. These components work in concert to support flexible exploration and precise filtering of the compound space.

\textbf{Attribute Distribution}\label{section::attribute distribution} (Fig. \ref{fig:teaser}\inlineicon{case_2_icons/A2.png}) integrates parallel coordinates plot and uncertainty visualization to reveal attribute relationships and prediction confidence. The visualization comprises two integrated parts:
(1) A parallel coordinates plot that reveals distribution patterns and relationships among attributes. Each axis represents one attribute, and polylines connecting these axes illustrate potential correlations between attributes (\textbf{T4}). Users can interactively brush along any axis to define specific value ranges, dynamically updating the range filtering criteria (\textbf{T5}). To facilitate efficient exploration, we group attributes based on their physical representations (e.g., bandgap-related attributes like gap, gap\_gw, and gap\_hse are grouped together). Additionally, to maintain system responsiveness with large datasets, we adopted a density-based sampling strategy, which will be further discussed in section \ref{section::DiscoveryView}. 
(2) An embedded uncertainty visualization using scatter plots between adjacent axes (Fig. \ref{fig:teaser}\inlineicon{case_2_icons/i.png}). Each point in these plots represents a compound's uncertainty values (discussed in section \ref{sec::explainable predictions}) for the corresponding attribute pair, where the distance from the origin indicates uncertainty; points closer to the origin suggest higher model confidence. This uncertainty representation enables users to assess prediction reliability during their filtering process (\textbf{T2}).

\textbf{The Exploration History} (Fig.~\ref{fig:teaser}\inlineicon{case_2_icons/A3.png}) supports tracking and navigation of previously applied filtering states (\textbf{T5}) using a tree diagram. 
Each filtering step is visualized as a transition, with edge color encoding the number of compounds retained. The system enables three filtering strategies: 
range filtering in the Filter View, and cluster and reference filtering in the Discovery View. Distinct icons represent range filtering (\inlineicon{visualdesign_icons/range_filter.png}) and cluster/reference filtering (\inlineicon{visualdesign_icons/cluster_filter.png}).

To facilitate statistical assessment, Analysis of
Variance (ANOVA) is integrated to determine the attribute with the highest variation when applying cluster and reference filtering, ensuring consistency across the three filtering strategies.
Hover interactions on the tree diagram highlight the corresponding attribute distributions in the Attribute Distribution view, enabling users to compare distributions across different filtering stages and better understand the progressive refinement process.

\subsection{Discovery View}
\label{section::DiscoveryView}
The Discovery View (Fig.~\ref{fig:teaser}\inlineicon{case_2_icons/B.png}) presents an overview of compounds within the current compound space, highlighting performance on user-selected key attributes. The compound space integrates original, AE-predicted, and CAMI-imputed values, focusing on similarity patterns, 
while detailed
compound attribute comparison is addressed in the Comparison View (Sec.~\ref{sec::comparison view}). This supports pattern exploration (\textbf{T3, T4}) and the identification of key attributes and their influencing factors (\textbf{T3}). A novel glyph-based representation is proposed to encode both key attribute values and influencing factors, and a new cluster-aware layout optimization technique is developed to reduce visual clutter and preserve the clustering of glyphs.
\par
\textbf{Glyph Design :}
 Inspired by matExplorer~\cite{matexplorer_pu2021}, we also employed a glyph-based representation (Fig. \ref{fig:teaser}\inlineicon{case_2_icons/B2.png}) to visualize each compound in the Discovery View. The glyph design consists of two integrated components: an inner pie chart and an outer circular bar chart, providing a multi-faceted visualization of compound attributes. Unlike matExplorer~\cite{matexplorer_pu2021}, which primarily focuses on the comparison and analysis of different ML models, our design is centered around supporting and facilitating compound discovery, particularly in the context of identifying promising semiconductors. Our glyph aims to make key property's influencing factors interpretable to the user, enabling them to actively explore, compare, and filter compounds based on domain-relevant criteria (\textbf{T3}).
 
\begin{figure}[tb]
    \centering
    \includegraphics[width=0.9\linewidth]{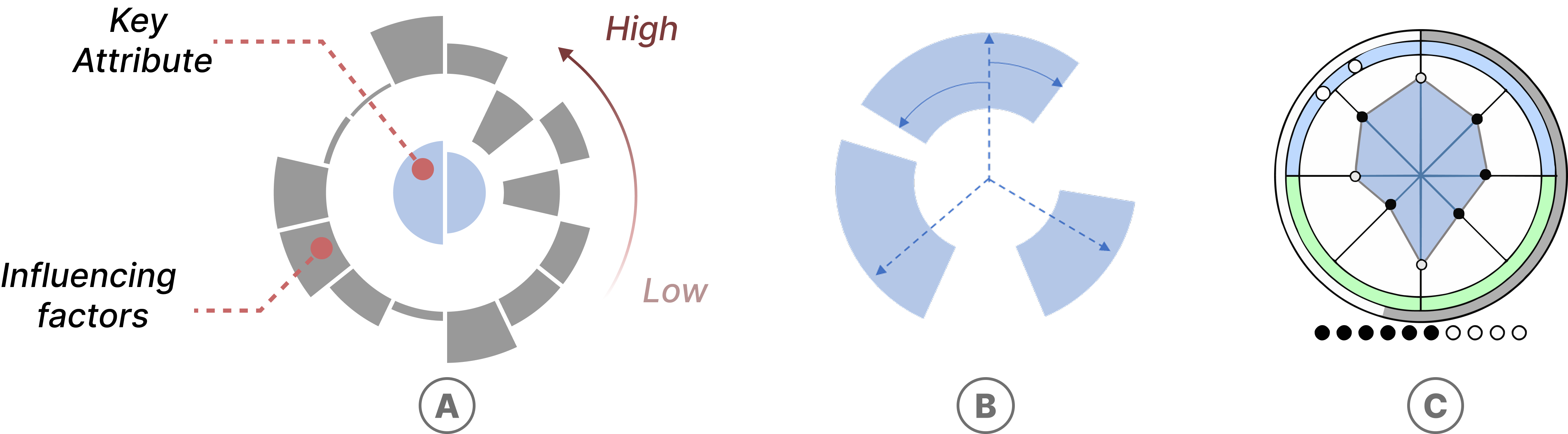}
    \caption{Design alternatives: (A) current glyph design; (B)  a circular bar chart design; (C) a radar chart design.}
    \label{fig:alternative designs}
\end{figure}

 As illustrated in Fig.~\ref{fig:alternative designs}A, the inner pie chart encodes the user-selected key attributes from the Control Panel. Each segment corresponds to one attribute, where segment size reflects its value. The segment color indicates the compound’s cluster membership, with distinct colors representing different clusters.

The outer circular bar chart extends this representation by encoding influencing factors through a systematic radial layout. In this way, glyph similarity and dissimilarity reveal not only how compounds compare in terms of values, but also the underlying reasons for these differences, thereby guiding users toward selecting further filtering attributes for more refined exploration. For each pie sector, we show the k most significant attributes ($k = \left\lfloor 15/n \right\rfloor$), where n is the number of sectors. The value 15 was empirically determined through iterative design and expert feedback to balance coverage and readability. The attributes are selected based on feature importance or correlation strength (\textbf{T3}).
For model-predicted attributes, we use SHAP values to reflect feature importance, while for other attributes, we use Pearson correlation coefficients with the inner key attributes to indicate correlation strength. To ensure a consistent and comparable representation, we use a unified encoding for these influencing factors in the outer circular bars, regardless of whether they originate from prediction or imputation. This avoids extra visual complexity in the glyph, and the distinctions in data source (Predicted, Imputed or Retrieved) are clarified in the Comparison View (Sec. \ref{sec::comparison view}). 
The attributes are ordered by decreasing significance counterclockwise, bar height encodes normalized values, and bar orientation represents relationship direction (inward for negative, outward for positive)(Fig. \ref{fig:alternative designs}A). The attributes influencing multiple inner attributes will appear accordingly in each relevant sector.

The glyph design allows users to focus on the selected attributes and also allows users to see the relationship between the attributes. Additionally, a tooltip showing the exact value and box plot of the selected attribute is displayed when users hover over a specific glyph, enabling detailed attribute exploration and comparison.

\textbf{Alternative Designs.}
We explored several alternative glyph representations before finalizing our design. The first approach (Fig. \ref{fig:alternative designs}B) used a circular bar chart with opposing bars along the circumference to show paired attributes like electron and hole mobility. While effective for highlighting relationships between paired properties, it became cluttered with more than three pairs and made cross-compound comparisons difficult. Our second design (Fig. \ref{fig:alternative designs}C) employed an enhanced radar chart with multiple encoding channels, including vertices for attribute values and black/white dots for ground truth versus predicted values. This could display more information but required significant cognitive effort to interpret. These explorations led to our current dual-layer design, which separates primary attributes (inner pie chart) from their relationships (outer circular bars), balancing visual clarity with analytical capability.

\par

\textbf{Cluster-Aware Layout Optimization.}\label{sec:cluster-aware-layout-optimization}
A major challenge in the Discovery View is the visual clutter created by overlapping glyphs, especially in dense areas with many compounds. This is because projecting key attributes and influencing factors into a 2D t-SNE embedding results in high-density clusters, as many compounds are closely related to each other. To address this, we use two strategies: density-based sampling and cluster-aware overlap removal.

\textbf{Density-Based Sampling.} We employ Kernel Density Estimation (KDE) to perform adaptive sampling while preserving the underlying data distribution. Based on the 2D embeddings obtained from t-SNE, we apply DBSCAN to identify clusters used for sampling. For clusters with more than three compounds, we compute a sampling rate proportional to their probability density

\(m_i = \max\left(3,\ \left\lfloor N \cdot \frac{n_i}{\sum_{k=1}^{K} n_k} \right\rceil\right), \quad 
\mathbb{P}(x_{ij}\ \text{is kept}) \propto \hat{f}_i(\mathbf{p}_{ij})\)

where $m_i$ is the number of samples kept from cluster $i$, $n_i$ is the number of compounds in cluster $i$, $K$ is the total number of clusters, and $N$ is the target sample size (set to 100 for optimal visualization). $\mathbb{P}(x_{ij}\ \text{is kept})$ denotes the probability that compound $x_{ij}$ is kept as a sample, $\mathbf{p}_{ij}$ represents the 2D embedding of compound $x_{ij}$ obtained by reducing the selected key attributes and corresponding influencing factors using t-SNE, and $\hat{f}_i$ is the kernel density estimate computed over the 2D embeddings of compounds in cluster $i$, using a Gaussian kernel with a bandwidth of 0.5 to balance local detail and smoothness in density estimation. This approach ensures that small clusters maintain structural integrity while larger clusters are sampled proportionally to density distribution so that the overall data distribution is preserved.

\textbf{Cluster-Aware Overlap Removal.} While overlap removal algorithms like DGrid, SORBID~\cite{hilasaca2024GridBased,giovannangeli2024Overlap} offer systematic solutions, they usually ignore the cluster structure within the data which is essential for material analysis. Cluster-aware methods~\cite{zhou2024ClusterAware}  addressed this limitation in grid-based layouts, however gridifying the scatter plot can significantly distort the original spatial relationships. To overcome these limitations, we propose a two-step overlap removal approach that maintains cluster integrity while mitigating glyph overlaps.

1) \textit{Space Optimization:} 
The first step transforms discrete point clusters into continuous spatial regions while preserving cluster relationships. Specifically, each cluster is represented as a convex hull polygon $P_i$, computed using the Quickhull algorithm~\cite{barber1996quickhull} on 2D embedding space. For the clusters with fewer than three points, we instead generate a circular buffer of fixed radius around the centroid to avoid degenerate polygons. Because clustering algorithms produce non-adjacent cluster polygons, dummy polygons are introduced to fill empty spaces for subsequent cartogram processing. These dummy regions are generated through a kd-tree inspired recursive partitioning approach. The global bounding box is initially divided into a $4\times4$ grid, and cells overlapping cluster hulls are recursively split along their longer axis until their area is below 0.1 (normalized units). The resulting regions are clipped to empty spaces, balancing computational efficiency and geometric fidelity.
With this continuous layout in place, we then employ the cartogram technique\cite{gastner2004diffusion} to optimize polygon areas. Each cluster polygon is scaled proportional to its number of member compounds, while dummy polygons are assigned a weight of 0.1 to act only as spacers. This ensures sufficient space for expansion and approximately preserves relative topological relationships.

2) \textit{Layout Refinement:} Within each optimized polygon, we use a force-directed algorithm to adjust compounds' positions. The total force $\mathbf{F}$ acting on each compound is defined as
$
\mathbf{F} = \mathbf{F}_b + \mathbf{F}_o + \mathbf{F}_{{l}} + \mathbf{F}_{{mb}} + \mathbf{F}_{{c}},
$
where $\mathbf{F}_{l}$, $\mathbf{F}_{mb}$, and $\mathbf{F}_{c}$ are standard forces from D3's force-directed simulation, handling edge attractions, node repulsion, and collision avoidance, respectively. Here, edges are defined within each cluster by connecting every compound to its nearest neighbor in the 2D embedding space, which provides the basis for edge attractions. Additionally, we introduce two specialized forces to maintain cluster structure and spatial relationships:

\begin{itemize}
    \item \textit{Boundary Containment Force} ($\mathbf{F}_b$): This force confines compounds within their designated cluster boundaries. If a compound drifts outside its cluster polygon, it is pulled back toward the nearest boundary point, given by $
      \mathbf{F}_b = k_b \cdot \frac{\mathbf{p}_{\text{nearest}} - \mathbf{p}}{\|\mathbf{p}_{\text{nearest}} - \mathbf{p}\|},
    $
    where $\mathbf{p}_{\text{nearest}}$ is the nearest point on the polygon boundary to compound position $\mathbf{p}$, and $k_b(20)$ controls the force magnitude.

    \item \textit{Position Preservation Force} ($\mathbf{F}_o$): This force preserves spatial relationships by pulling compounds toward their initial positions, defined as
      $\mathbf{F}_o = k_o \cdot (\mathbf{p}_{\text{i}} - \mathbf{p}),$
    where $\mathbf{p}_{\text{i}}$ is the initial position and $k_o(0.2)$ is the attraction strength.
\end{itemize}
The layout optimization proceeds iteratively until reaching equilibrium, with force magnitudes adjusted dynamically to balance position preservation, overlap prevention and cluster integrity.

\subsection{Comparison View} \label{sec::comparison view}
Comparison View (Fig.\ref{fig:teaser}\inlineicon{case_2_icons/C.png}) facilitates detailed comparative analysis of candidate compounds (\textbf{T6}) through an interactive heatmap visualization. 
For each candidate compound, we use a heatmap (Fig. \ref{fig:teaser}\inlineicon{case_2_icons/C1.png}) to visualize the values of each attribute, with rows representing candidate compounds and columns representing attributes. The heatmap is color-coded to indicate the value of each attribute, with a gradient from light yellow (low) to blue (high), enabling rapid identification of patterns and outliers. Rows are ordered according to chemical formula similarity, ensuring that compounds with related compositions are positioned adjacently, while columns are grouped by predefined categories as described in section \ref{section::FilterView}. Additionally, shapes encode the data source: circles \inlineicon{visualdesign_icons/heatmap_circle.png} indicate predicted values (AE \& NGB), triangles \inlineicon{visualdesign_icons/heatmap_triangle.png} represent imputed values (CAMI), and rectangles \inlineicon{visualdesign_icons/heatmap_rectangle.png} denote ground truth. For predicted values, circle radius reflects uncertainty, with larger radius indicating lower uncertainty. To enhance analytical precision, a tooltip reveals the exact attribute value upon hovering over the cell.

For crystal structure, a non-numerical compound characteristic, \name\ integrates interactive 3D visualizations using 3DMol.js (Fig.~\ref{fig:teaser}\inlineicon{case_2_icons/C2.png}). Hovering over the heatmap opens an interactive 3D molecular viewer with rotation, zooming, and screenshot capture interactions enabled, which helps users intuitively explore compound structures.


\section{Evaluation}
We evaluate our approach using both quantitative and qualitative approaches. The following section discusses the findings.

\subsection{Quantitative Evaluation} \label{sec:quantitative-evaluation}
 We conducted two quantitative evaluations to validate model selection and assess the proposed imputation method.

\textbf{Model Evaluation.} We evaluated eight ML models, commonly used in material science, for predicting five semiconductor attributes using MAPE, including RF, SVR, LR, MLP, DT, GB, NGBoost, and a TensorFlow-based fine-tuned AE (See Appendix
A.4 for architecture). As the dataset was small, we used leave-one-out cross-validation with three validation samples and repeated all experiments five times with different random seeds for robustness.

As shown in Table \ref{tab:model_comparison}, the AE model performs best for carrier mobility, whereas NGBoost is superior for bandgap GW, VBM GW, and CBM GW predictions. These differences likely reflect the limited data availability described in Section \ref{section::Intro}.

\begin{table}[htbp]
    \centering
    \begin{tabular}{lrrrrrr}
        \toprule
        Model & E Mob & H Mob & Bandgap & VBM & CBM  \\
        \midrule
        RF & 109.00 & 203.38 & 7.25 & 2.71 & 6.29\\
        SVR & 104.25 & 242.39 & 41.53 & 4.74 & 5.68 \\
        LR & 321.60 & 1,779.28 & 21.19 & 8.48 & 19.28 \\
        MLP & 178.90 & 195.58 & 10.95 & 3.74 & 7.81\\
        DT & 122.35 & 186.01 & 9.49 & 3.71 & 8.24 \\
       GB & 88.34 & 172.31 & 5.83 & 2.35 & 5.36 \\
        NGBoost & 90.22 & 170.98 & \textbf{5.74} & \textbf{2.30} & \textbf{5.34}  \\
        AE & \textbf{71.39} & \textbf{102.70} & 27.17 & 16.95 & 24.75 \\
        \bottomrule
    \end{tabular}
    \caption{Comparison of MAPE for the evaluated models across five semiconductor properties. ``E Mob'' and ``H Mob'' refer to electron and hole mobility, respectively, while bandgap, VBM, and CBM use GW-calculated ground truth values.}
    \label{tab:model_comparison}
\end{table}

\textbf{Imputation Method Evaluation.}
We evaluate CAMI against the most popular imputation methods in materials science domain, such as mean imputation and zero imputation. These were selected as they were \textbf{commonly used in materials science workflows}~\cite{ma_mastery_2024_imputation,dunn_benchmarking_2020_imputation}. 
Therefore, other multivariate imputations were not included in this comparison. According to table \ref{tab:imputation_comparison},  
CAMI has a significantly higher performance in E Mobility and Hole mobility prediction while it only shows some improvement over mean imputation for Bandgap GW VBM GW, CBM GW. The evaluation was conducted by comparing the prediction performance of the AEs for electron and hole mobility, and NGBoost for the other properties. We speculate this is a result of not filling all the missing values with the same value, which preserve the correlation between attributes of the compounds, especially when there are few ground truth data available.

\begin{table}[htbp]
    \centering
    \begin{tabular}{l r r r r r r}
        \toprule
        \textbf{Method} & \multicolumn{5}{c}{\textbf{Desirable Semiconductor Attribute}}\\
        \cmidrule(lr){2-6}
        & \textbf{E Mob} & \textbf{H Mob} & \textbf{Bandgap} & \textbf{VBM} & \textbf{CBM} & \\
        \midrule
        Mean           & 71.39 & 102.70 & 5.74 & 2.30 & 5.34 \\
        Zero          & 68.96 & 116.78 & 7.19 & 2.68 & 5.36 \\
        MF & 70.65 & 120.11 & 7.19 & 2.68 & 5.36 \\
        CAMI & \textbf{63.53} & \textbf{98.17}  & \textbf{5.67} & \textbf{2.28} & \textbf{5.26}  \\
        \bottomrule
    \end{tabular}
    \caption{Comparison of imputation methods. E Mob and H Mob (electron and hole mobility) are predicted with AE, while Bandgap, VBM, and CBM are predicted with NGBoost, using GW-calculated values as ground truth. MF denotes Most Frequent.}
    \label{tab:imputation_comparison}
\end{table}

\subsection{Expert Interview}
We conducted semi-structured interviews with 12 experts (Male = 8) including 8 PhD students, 1 postdoctoral researcher and 3 research fellows, averaging 5.2 years of research and engineering experience in semiconductors
or other related materials.

\par
\begin{figure}[tb]
  \centering
  \includegraphics[width=\columnwidth]{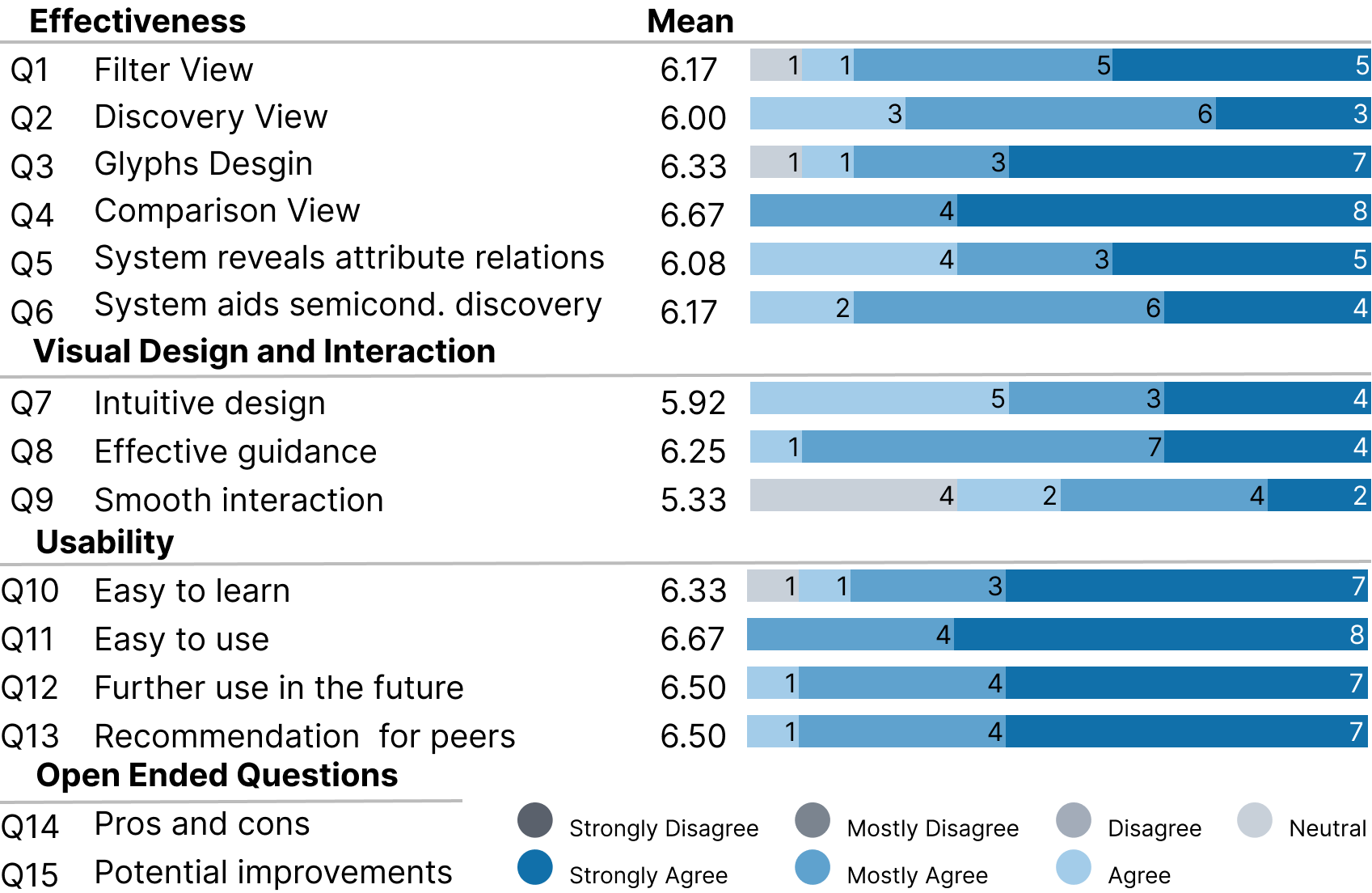} 
  \caption{The expert interview questionnaire results.}
  \label{fig:results} 
\end{figure}

\textbf{Procedures:} Each expert interview (approx. 60 minutes) comprised three phases: (1) a system introduction covering the research background, visualization design, system interactions, and a 15-minute usage scenario; (2) two assigned tasks (approx. 30 minutes); and (3) a post-study questionnaire (Q1-Q15), as shown in Fig. \ref{fig:results}. Q1-Q13 used a 7-point Likert scale adapted from PSSUQ~\cite{pssuq_lewis_psychometric_1992} and prior work~\cite{ponzilens_wen_ponzilens_2024} to evaluate effectiveness (Q1-Q6), design and interaction (Q7-Q9), and usability (Q10-Q13). Q14-Q15 collected open-ended feedback. Participants received approximately 15 USD as compensation.

\subsubsection{Results}

The results are summarized in Fig. \ref{fig:results}, with full questions in Appendix A.5. Overall, \name\ received high ratings, with no negative scores (1-3 on the Likert scale). Neutral ratings were given for Q1, Q3, Q9, and Q10 (1, 1, 4, and 1 respectively), notably by E7 for Q1 and Q3.

\textbf{Effectiveness.} According to scores from Q1–Q6, all but one of the participants agreed that the system is an effective tool for 2D semiconductor discovery. This is reflected in consistently high mean scores ranging from 6.00 to 6.67.

In terms of qualitative feedback, E7 rated neutral for Q1 and Q3, stating the main reason as ``the amount of information displayed is overwhelming''. They further elaborated that the ``text in the Filter View is too small'' and that they ``would like a feature where I can toggle some of the view off to get more screen space for the view I’m interested in''. They also suggested to ``add a tool tip for supporting attributes in addition to the glyph legend''.

Other experts emphasized effectiveness-related aspects, agreeing that the Filter View is ``useful for the initial stage of material selection'' and that the Discovery View can be utilized to ``easily find related materials and alternative materials'' and that the glyph design supports filtering by linking property values with their influencing factors. E7 highlighted that the filter history view is useful, while E11 and E12 appreciated the predicted values of the system, which mitigate the necessity of ``hectic experimental procedures''.

\textbf{Visual Design and Interaction.} Most participants agreed that the system has a good visual design and interaction quality, with mean scores between 5.33 and 6.25. 
This indicates general satisfaction.Qualitative feedback confirmed this. Most participants agreed the system is ``visually appealing'', which is consistent with quantitative ratings. Meanwhile, participants also provided suggestions for further improvement. For example, E4, E7, E8, and E11 provided a neutral score for Q9, stating that the system speed needs to be improved, as it takes about 7 seconds to update the Discovery View when there are a substantial number of elements due to computationally intensive cartogram calculations. E11 also stated that ``the system needs more explanations such as more detailed labels and legends'' and rated a neutral score for Q12. Other experts suggested improvements to the user interface, such as ``providing more detailed legends and information panels'' and ``using larger font size and high-contrast colors''.

\textbf{Usability.} According to scores from Q11–Q13, all participants agreed that the usability of the system is excellent, with a mean rating of 6.50. This represents the highest overall score among the three categories.
Despite the issues mentioned in the visual design and interaction section, all the experts agreed that the system is easy to use and that they would like to use it to identify potential semiconductors in the future. They also agreed that they would recommend this system to other material scientists.

\subsection{Use Cases}

\textbf{Case 1: Exploring clusters - Identifying 2D Materials for Water Splitting}\par

\begin{figure}[tb]
  \centering
    \setlength{\abovecaptionskip}{0.2cm}
  \setlength{\belowcaptionskip}{-0.4cm}
  \includegraphics[width=\columnwidth]{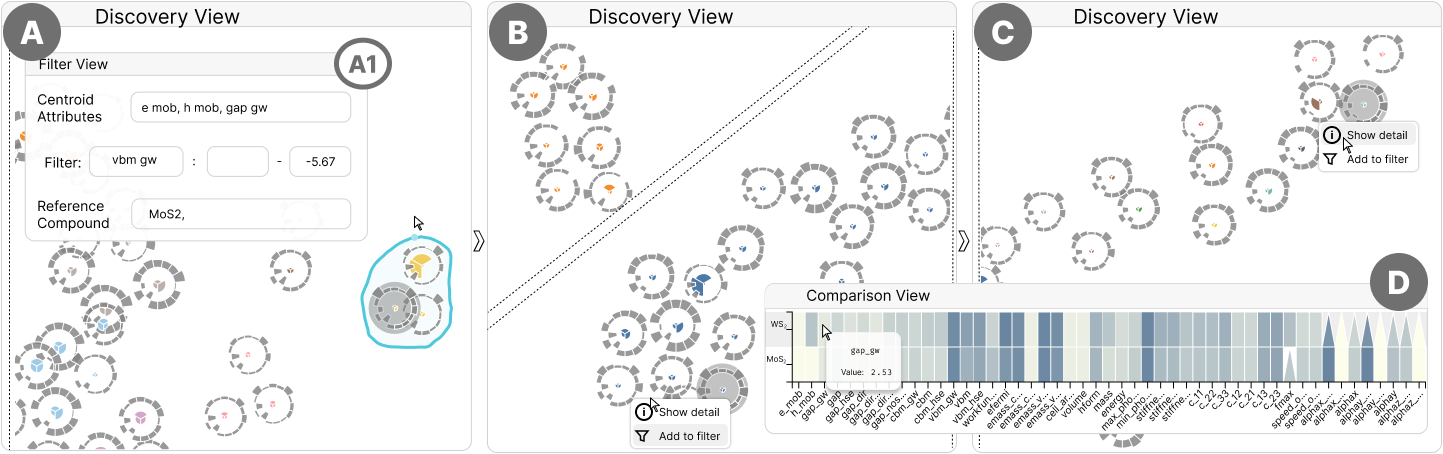} 
  \caption{ 
  With \name\ an expert has identified WS$_2$ as a similar material to MoS$_2$, suitable for hydrogen evolution reaction in water splitting. First, they applied 2 filters, CBM GW $>$ -4.44eV and VBM GW $<$ -5.67 eV using the Filter View. Then they selected MoS$_2$ as the reference compound (\inlineicon{case_study_1_icons/A1.png}). From the resulting Discovery View, they identified the cluster with MoS$_2$ and selected it by lassoing the cluster (\inlineicon{case_study_1_icons/A.png}). They identified the MoS$_2$ from the resulting Discovery View (\inlineicon{case_study_1_icons/B.png}) and add it as the filter for next step. Then they identified WS$_2$ as the nearest neighbor for MoS$_2$ from the Discovery View (\inlineicon{case_study_1_icons/C.png}) and added both to comparison view for further exploration (\inlineicon{case_study_1_icons/D.png}). Some compounds are omitted from the Discovery View, marked by dotted lines. See Appendix A.9 for the high resolution version.
  \label{fig:case_study_1} 
  }
\end{figure}

Photocatalytic water splitting refers to the light-driven decomposition of water into H$_2$ and O$_2$ using semiconductor photocatalysts, which can be achieved either in a single-step system with suitable band alignment or in a Z-scheme employing two semiconductors for separate reduction and oxidation~\cite{yang_recent_2020_water_splitting}.

In this use case, we demonstrate how \name{} can be used
to identify 2D semiconductors for photocatalytic water splitting. Effective splitting requires \textbf{proper CBM and VBM alignment with water redox potentials}, i.e., the CBM GW must lie above –4.44 eV for H$^+$/H$_2$ reduction, and the VBM GW below –5.67 eV for O$_2$/H$_2$O oxidation~\cite{wang_ml-aided_2024_water_splitting_CBM_VBM}. These criteria served as the initial filtering parameters (Fig. \ref{fig:case_study_1} \inlineicon{case_study_1_icons/A.png}). Recognizing MoS$_2$ as a benchmark catalyst for Hydrogen Evolution Reaction (HER)~\cite{yu_advanced_2023_MoS2_WS2_water_splitting}, the user set it as a reference in the Filter View (Fig. \ref{fig:case_study_1} \inlineicon{case_study_1_icons/A1.png}). Observing a cluster of compounds around MoS$_2$ (Fig. \ref{fig:case_study_1} \inlineicon{case_study_1_icons/A.png}), the user lassoed this region to isolate candidates (Fig. \ref{fig:case_study_1} \inlineicon{case_study_1_icons/B.png}). To further refine the results, MoS$_2$ was selected as the reference material, yielding the 30 closest compounds (Fig. \ref{fig:case_study_1} \inlineicon{case_study_1_icons/C.png})~\textbf{(T5)}.

For comparative evaluation, MoS$_2$ and its nearest neighbor WS$_2$ were selected (Fig. \ref{fig:case_study_1} \inlineicon{case_study_1_icons/C.png}) \textbf{(T6)}. The user observed that their bandgaps (2.53 eV) slightly exceeded the ideal 1.5 to 2.5 eV range~\cite{zuo_review_2024_water_splitting_bandgap}, which balances visible light absorption with enough photovoltage to drive H$_2$/O$_2$ evolution, for visible-light photocatalysis but remained acceptable (Fig. \ref{fig:case_study_1} \inlineicon{case_study_1_icons/D.png}). External validation also confirmed WS$_2$ as a promising HER catalyst~\cite{yu_advanced_2023_MoS2_WS2_water_splitting}, highlighting the system’s effectiveness in rapidly identifying candidate 2D semiconductors for water splitting.

\par
\textbf{Case 2: Sequential Filtering - Discovering 2D Materials for Tunnel Field-Effect Transistors (TFETs)}\par

A TFET is an electronic switch that leverages quantum tunneling rather than conventional current flow to enable lower-power operation than traditional transistors. TFETs require \textbf{``Type-II band alignment''}, where \textbf{the CBM of one material lies below the VBM of another to enable tunneling}~\cite{TFET_verma_comparative_2024}. This case shows how \name{} uses predicted CBM and VBM GW values to identify 2D donor–acceptor pairs for TFETs.
First, the user selected \textit{gap\_gw,vbm\_gw and cbm\_gw} as the centroid attributes \textbf{(T3)}. Initially, the user defined TFET-specific filters, selecting semiconductors with bandgaps of 0.5 to 2.0 eV, since bandgaps below 0.5 eV prevent turn-off and those above 2.0 eV hinder tunneling efficiency~ (Fig.\ref{fig:teaser}\inlineicon{case_2_icons/A1.png})\cite{klinkert_ab_2018_TFET_bandgap}. Donor candidates were further filtered by low CBM value (typically around -4.5 eV or lower) using the control panel in Filter View (Fig. \ref{fig:teaser}\inlineicon{case_2_icons/A1.png})~\textbf{(T5)}. Recognizing SnSe$_2$ as a promising donor candidate~\cite{SnSe2_sato_intrinsic_2021} based on prior domain knowledge, the user selected it for further analysis (Fig.\ref{fig:teaser}\inlineicon{case_2_icons/A2.png}). In the Discovery View, SnSe$_2$ was found to have a \textbf{VBM of -6.88 eV and a CBM of -4.97 eV (Fig. \ref{fig:teaser}\inlineicon{case_2_icons/B3.png})(T4)}.

To identify acceptors forming a Type-II heterostructure with SnSe$_2$, the user removed the CBM filter and applied band alignment criteria of \textbf{CBM GW lower than -4.97 eV and VBM GW higher than -6.88 eV.} Using the Attribute Distribution view (Fig. \ref{fig:teaser}\inlineicon{case_2_icons/A2.png}), candidates were further constrained to energy above hull values below 0.1 eV for thermodynamic stability and electron mobilities above 200 cm$^2$V$^{-1}$s$^{-1}$ for high performance suitability \textbf{(T5)}.
These criteria yields 11 potential candidates (Fig.\ref{fig:teaser}\inlineicon{case_2_icons/C1.png}): AgTlP$_2$S$_6$, AgTlP$_2$Se$_6$, Au$_4$I$_8$, Cu$_4$I$_4$S$_4$, Cu$_4$I$_4$Se$_4$, Ir$_2$Nb$_2$S$_8$, PbBi$_2$Se$_4$, PbSb$_2$Se$_4$, Rh$_2$Ta$_2$S$_8$, Rh$_4$S$_8$, and SnBi$_2$S$_4$. The user was able to identify the confidence of the predicted electron mobility values that affect the performance, as the uncertainty for each predicted value was shown (Fig.\ref{fig:teaser}\inlineicon{case_2_icons/C1.png}) \textbf{(T2)}. The user used domain knowledge for further refinement. For instance, Au$_4$I$_8$ (Fig.\ref{fig:teaser}\inlineicon{case_2_icons/B1.png}\inlineicon{case_2_icons/C2.png}) was excluded as a non-layered structure, unsuitable for high-quality van der Waals interfaces in TFETs. This domain-informed filtering shows the effectiveness of combining computational insights with expert knowledge to identify viable 2D semiconductors.

\section{Discussion}
In this section, we discuss lessons learned during the development and evaluation of \name{}, as well as its possible limitations.

\textbf{Generalizability.} Although \name\ uses C2DB data and targets 2D semiconductor discovery, its workflow, including CAMI, is database-independent and can be applied to other datasets. For instance, the Materials Project database could support bulk material discovery, but ML models and CAMI must be retrained with appropriate hyperparameter tuning.

\textbf{Human-AI collaboration for new material discovery.} As noted in the introduction, human intuition plays a vital role in the discovery of new materials. \name\ exemplifies how domain experts can harness the strengths of ML models to facilitate the material discovery process.

\textbf{System speed and user perception.} The system exhibited an average response delay of around 7 seconds due to the iterative overlap removal process (Section \ref{section::DiscoveryView}) and the complexity of high-dimensional datasets. Our participants rated Q9 lowest (mean 5.33), which reflects the relatively slow response. Response speed can be further improved, which is left for future work.

\section{Conclusion and Future Work}
We present \name, a visual analytics system to support discovery of new 2D semiconductors. Eight ML models were evaluated to identify suitable predictors for five hard-to-compute attributes, and a correlation- and similarity-based imputation method was proposed to handle missing values. Integrated into \name, these components form a comprehensive workflow. Two use cases and interviews with 12 material science experts evaluated effectiveness, design, interaction, and usability. Findings show that \name\ effectively supports semiconductor discovery.

Future work will extend \name\ to additional databases beyond 2D materials and experimentally validate shortlisted compounds in collaboration with material scientists.
\par
\textbf{Acknowledgements}
We used ChatGPT solely for grammar correction and language polishing.
This project is supported by the Ministry of Education, Singapore, under its Academic Research Fund Tier 2 grant (T2EP202220049), Tier 1 grant (22-SIS-SMU-054), and NTU Start Up Grant awarded to Yong Wang.


\clearpage
\bibliographystyle{eg-alpha-doi} 
\bibliography{src/template}       


\end{document}